\RequirePackage{fix-cm}
\documentclass[smallextended]{svjour3}       
\smartqed  
\usepackage{bm}
\usepackage{graphicx}
\usepackage{lineno}
\begin{document}

\title{Elastic characteristics for naturally-fractured reservoirs using an integrated LSM-DFN scheme with rough contact deformation
}

\titlerunning{Elastic characteristics for naturally-fractured reservoirs}        

\author{Ning Liu
}


\institute{N. Liu \at
College of Mechanical and Electrical Engineering, Beijing University of Chemical Technology, Beijing 100029, China\\
              \email{nicola1sunshine@gmail.com}           
}

\date{Received: date / Accepted: date}

\maketitle

\begin{abstract}
Crack micro-geometries and tribological properties pose an important impact on the elastic characteristics of fractured rocks. Numerical simulation as a promising way for this issue still faces some challenges. With the rapid development of computers and computational techniques, discrete-based numerical approaches with desirable properties have been increasingly developed, but few attempts to consider the particle surface roughness in a lattice type model. For this purpose, an integrated numerical scheme accounting rough contact deformation is developed by coupling modified LSM and DFN modeling for predicting the effective mechanical properties of a realistic outcrop. Smooth joint logic is introduced to consider contact and slip behaviors at fracture surfaces and a modified contact relation to estimating the normal force-displacement from rough contact deformation. Improved constitutive laws are developed and employed for rock matrix and rough fracture surface and implemented in the modified LSM. Complex fracture networks presented by DFNs are automatically extracted based on the gradient Hough transform algorithm. This developed framework is validated by classic equivalent medium theories. It shows the model could be used to emulate naturally-fractured media and to quantitatively investigate the effects of fracture attributes and micro-scale surface roughness on the compression mechanism.
\end{abstract}

\section{Introduction}
\label{s1}
Naturally-fractured reservoirs have long been a significant target for the oil/gas industry \cite{grechka2006effective}. Fracture networks provide permeable conduits for fluid flow, which poses a paramount influence on the mechanical and transport properties. Therefore, assessing and understanding the effective mechanical properties of the fractured system is a prerequisite to many geotechnical applications and a still major scientific issue \cite{davy2018elastic}. The difficulties are mainly from the complexity of the distribution and density of the fractures: Fractures vary in multi-scales with the highly-variable density. This leads to the fact that the geometrical and statistical models of fracture patterns are still debated \cite{davy2018elastic}. Many theoretical models have been proposed to analytically describe the effective elasticity of constituent fractures \cite{bristow1960microcracks,cheng1993crack,eshelby1957determination,hudson1980overall,hudson1994overall}, but limited to the idealization and oversimplification of the naturally-fractured media \cite{liu2020stress}. As an efficient alternative for general applicability, numerical methods are extensively applied to the effective elastic modeling of constituent fractures. 

Numerical simulations as an efficient supplement to experimental measurements, provide an independent verification of theoretical predictions. Generally, numerical approaches are divided into two categories: Continuum methods and discrete-based models \cite{liu2020modified}. The former is based on a governing equation of stress and strain that is formulated under the frame of continuum mechanics \cite{liu2020modified}. The latter regards rocks as an assembly of microstructural elements that interact with each other by microstructural forces \cite{liu2020modified}. For the numerical simulation of naturally-fractured reservoirs, it mainly raises two main issues \cite{nadimi2019effect}. The first issue is still the intrinsic complexity of fracture networks \cite{bonnet2001scaling,davy2018elastic}. The other one is about the simulations of the various behaviors at the fracture surfaces, e.g., stick and slip behaviors, contact behaviors caused by surface roughness, or even fracture initiation, propagation, and closure process. 

Discrete fracture network (DFN) is developed to emulate complex fracture systems, referring to a geometry configuration of fractures \cite{liu2020stress}. As mentioned by N. Liu and Fu \cite{liu2020stress}, combined with continuum- and discrete-based approaches, the DFN is widely used in various engineering analyses with fractured rocks \cite{alghalandis2017similarity,brady1993rock,jing2003review,jing2007fundamentals}. Lei et al. \cite{lei2017use} combine continuum-based approaches with DFN could model fractured rocks with only a few cracks or plenty of fractures accounting a very small amount of displacement/rotation by introducing interface elements, or joint elements \cite{goodman1968model,lei2017use}. While, such a treatment is difficult to deal with stick and slip behaviors around fracture surface, complete detachment or large-scale fracture opening problems \cite{jing2003review,zhao20113d,lei2017use}, which are the key issues in the aforementioned. Modeling the high-density and complex DFNs remains difficult, which is regarded as the intrinsic limitation of continuum-based methods \cite{jiang2017crack}. For more complex DFNs, discrete-based approaches seem more suitable \cite{jing2002numerical}, especially for fractured rocks with a wide range of mineral compositions and fabric anisotropies. 

Discontinuum-based methods, like molecular dynamics (MD), lattice spring model (LSM) \cite{hrennikoff1941solution}, and discrete element method (DEM) \cite{cundall1971computer}, are commonly used to emulate mechanical deformations in rocks. Based on the approaches, inhomogeneous effects at the microlevel could be captured \cite{liu2020elastic,liu2020stress,liu2020modified,reck2017lattice,suiker2001comparison}, where granular textures, particle-scale kinematics, and force transmission can be correlated at the microlevel. Harthong et al. \cite{harthong2012strength} propose a coupled DEM-DFN model for characterizing the strength of rock masses. Among these discrete-based methods, the LSM attracts the most interest because it is flexible to model both continuous and discontinuous systems in a discrete way \cite{liu2020modified,22}. Unlike the DEM where elements interact through contact surfaces, the LSM connects elements by springs or beams, with several desirable properties, such as broad applicability, easy implementation, and high flexibility to handle the contact complexity of granular materials. Recently, more advanced LSMs are developed to avoid the Poisson's ratio limitation of the early LSMs. For example, N. Liu et al. \cite{liu2020modified} propose a modified LSM model by introducing an independent micro-rotational inertia to avoid the Poisson's ratio limitation and characterize the scale-dependent effects. Based on the modified LSM, N. Liu and Fu \cite{liu2020stress} develop a coupled LSM-DFN model to investigate the stress-orientation effect on the effective elastic anisotropy of complex fractures. The model is validated by Brazilian tests with different loading orientations, but has not yet been compared with classic theoretical models. There remain limitations in the model that cannot deal with stick and slip behaviors around fracture surfaces with rough contact deformation. Moreover, the DFN generation extracted manually \cite{liu2020stress} is quite time-consuming. 

In this study, we improve the coupled LSM-DFN model by introducing a smooth joint logic to consider contact and slip behaviors at fracture surfaces and a modified contact relation to estimate the normal force-displacement from rough contact. The DFNs are extracted automatically by an image process of naturally-fractured rocks based on the gradient Hough transform. The integrated LSM-DFN scheme is validated by classic equivalent medium theories, enabling the quantitative investigation for the effects of fracture attributes and micro-scale surface roughness on the elastic and anisotropic characteristics. The rest of the paper is organized as follows: First, we introduce the extraction and characterization of geometrical attributes of the target sample to indicate the complexity of natural fractures in Sect.\ref{s2}; Second, the sample is emulated by an improved LSM-DFN coupling model incorporating the surface roughness effects in Sect.\ref{s3}. Then, the implementation of the introduced constitutive laws and the integrated workflow of this LSM-DFN are verified by comparison with theoretical predictions in Sect.\ref{s4}; Section \ref{s5} presents the numerical results from LSM-DFN modeling by uniaxial compression tests to study how the realistic fracture attributes and micro-scale surface roughness affect the overall elasticity. The conclusions and future work are underlined in Sect.\ref{s6}.

\section{Fracture Extraction and Statistical Analysis}
\label{s2}
\subsection{Image Processing and Fracture Extraction}
\label{s2.2}
\begin{figure}
\centering\includegraphics[width=1.0\linewidth]{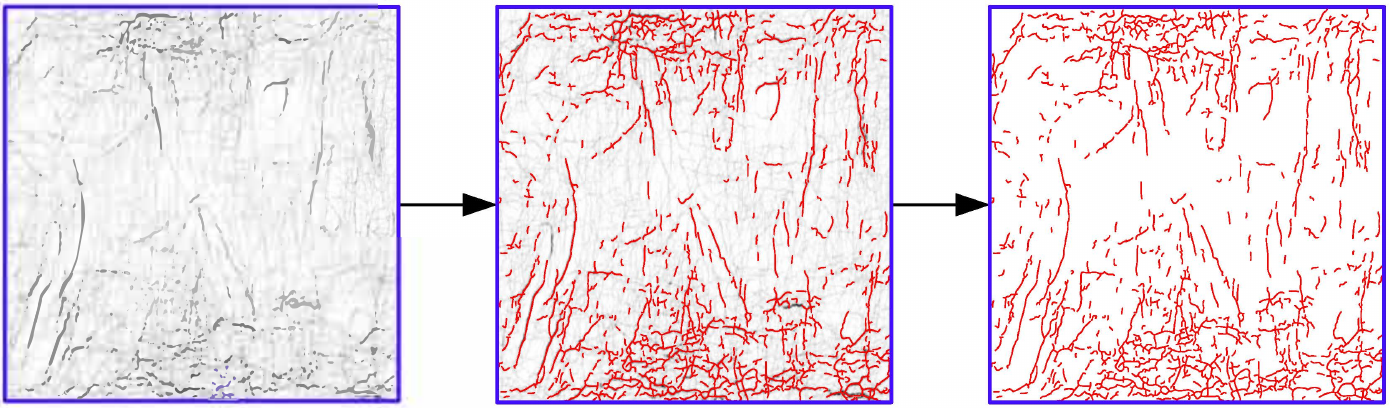}
\caption{Flow diagram for crack extraction from a digital photograph of an outcrop based on GrdHT algorithm}\label{f2}
\end{figure}

Reservoir rocks commonly contain pores, cracks, and fractures, exhibiting heterogeneity and anisotropy \cite{liu2020elastic}. Obtaining the geometrical attributes of the constituent fractures serves as a prerequisite for constructing realistic geological and numerical models for geo-mechanical and hydrological behavior studies. Complex fractured system as shown in Fig. \ref{f2} raises challenges in a high-accurate modeling for natural fracture geometry. N. Liu and Fu \cite{liu2020stress} extracted manually from a digital photograph by the instruction of Healy et al. \cite{healy2017fracpaq}. Manual extraction is the most widely used method, completely depending on the specialist's knowledge and experience with limited accuracy and expensive time-cost. Therefore, automatic image-based fracture extraction could be treated as an alternative replacement.

Extracted fractures could be treated as continuous lines are composed of some straight segments delimited by nodes \cite{liu2020stress}. Hough transform (HT) and inverse Hough transform are commonly used for the line detection and reconstruction \cite{hassanein2015survey}. Here, we use the gradient Hough transform (GrdHT) introduced by Hassaneinet al. \cite{hassanein2015survey} for efficient line detection. For simplicity, raw color images of outcrops captured by digital cameras are converted into grayscale representations of 8-bit binary images. The GrdHT algorithm calculates the gradient magnitudes for the image pixels. As shown in Fig. \ref{f2}, those with a gradient magnitude less than a certain threshold could be separated from the image background to segment the image and extract the crack networks. With different thresholds, smaller scale cracks could be captured. Compared with the fracture networks displayed in our previous work \cite{liu2020stress}, more and smaller scales are captured efficiently.

\subsection{Statistical Analysis of Crack Characteristics}
\label{s2.3}
\begin{figure}
\centering\includegraphics[width=1.0\linewidth]{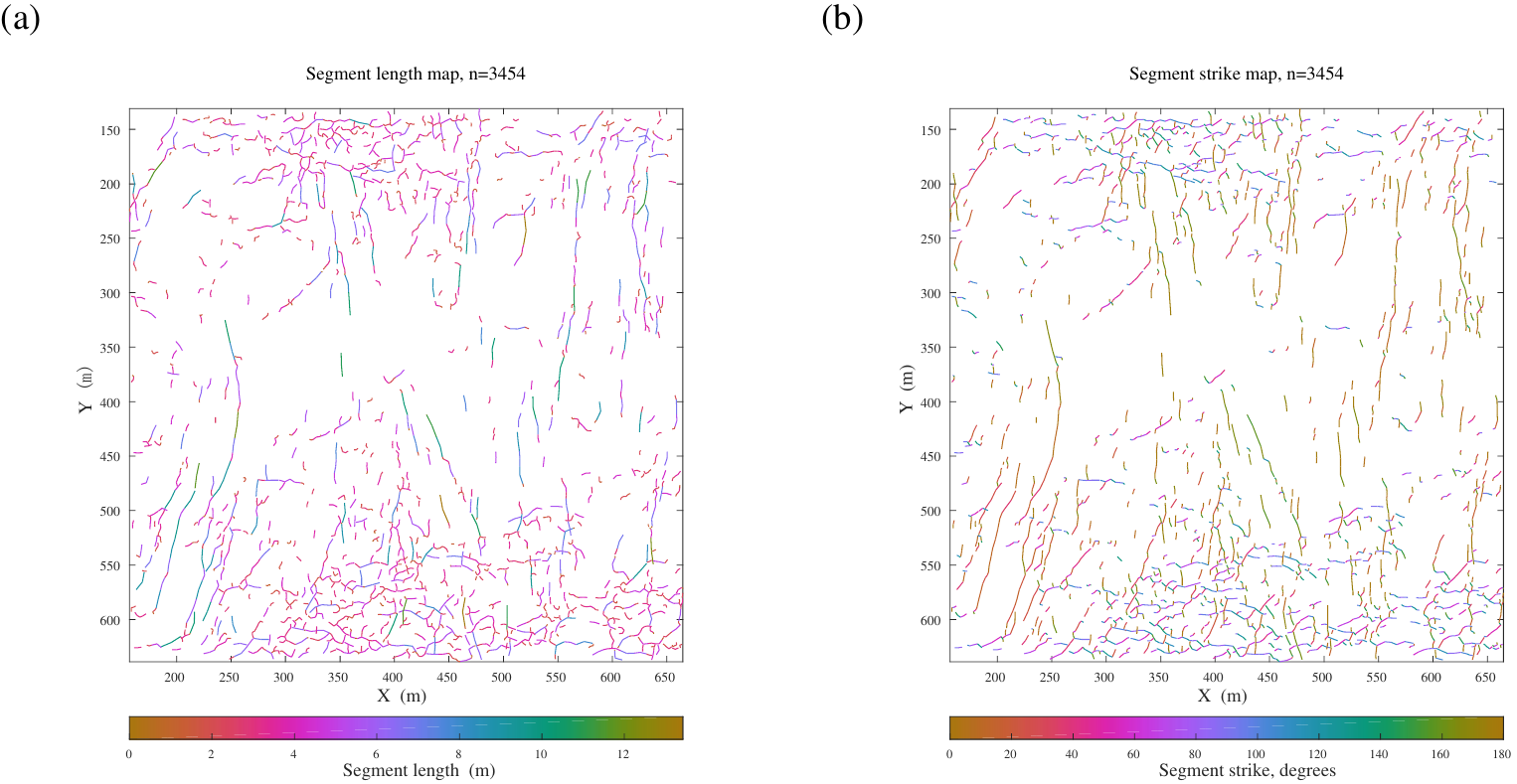}
\caption{Segment length (a) and orientation (b) distribution of the constituent segments}\label{f3}
\end{figure}

\begin{figure}
\centering\includegraphics[width=1.0\linewidth]{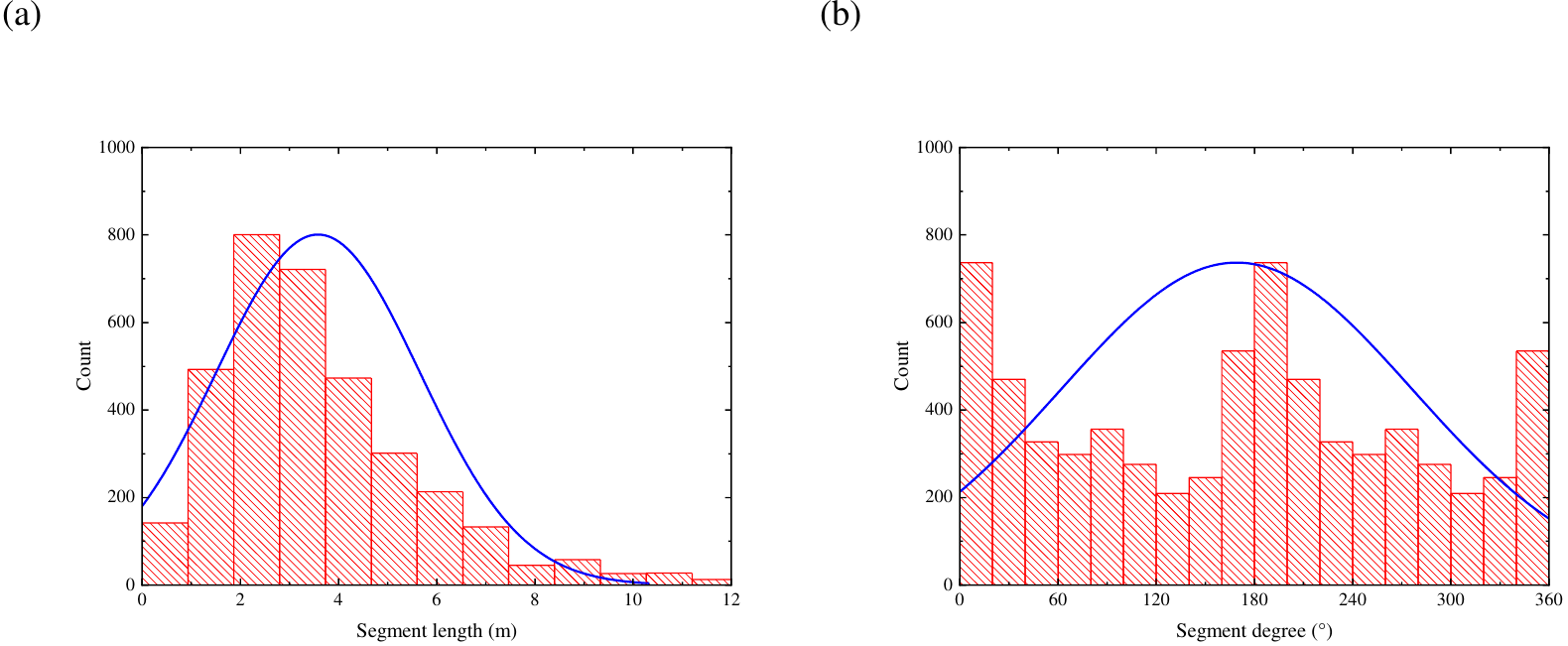}
\caption{Quantification for the lengths (a) and orientations (b) of the constituent segments}\label{f4}
\end{figure}

Fractures at any scale are believed to form through the interaction and coalescence of smaller fractures, namely fracture segments. Those fracture attributes and spatial variation are qualified by FracPaQ toolbox \cite{healy2017fracpaq} using simple operations in coordinate geometry. Figures \ref{f3} and \ref{f4} show length and orientation statistics of fracture segments. We could see the segment lengths vary from 0 to 14 m, mainly between 1$\sim$5 m, and the angles are distributed and concentrated within the range of ${\rm{ - 2}}{{\rm{0}}^ \circ } \sim {\rm{4}}{{\rm{0}}^ \circ }$, or ${\rm{16}}{0^ \circ } \sim {\rm{22}}{0^ \circ }$. The orientation distribution in a fracture pattern is important for unravelling the tectonic history of the rocks and in controls rock mass behavior with respect to attributes \cite{liu2020stress}.

\section{Integrated LSM-DFN Scheme}
\label{s3}
To build an LSM-DFN coupling model for the rocky outcrop incorporating the surface roughness effects, the rock matrix and fracture geometries are represented by lattice spring models (LSMs) and discrete fracture networks (DFNs), respectively. Smooth joint logic (SJL) accounts for the movements along fracture surface. A modified contact relation proposed by Greenwood and Tripp \cite{greenwood1967elastic} and extended by Nadimi et al. \cite{nadimi2019effect} is introduced to estimate the normal force-displacement from rough contact .

\subsection{Fracture Contact Model for Interfaces}
\label{s3.2}
Natural cracks may have irregular shapes and the fracture surfaces have roughness to some degree. When two rough fracture surfaces pressed against one another, the effective stiffness of the interfaces are known to be affected by surface roughness \cite{liu2005effects}. However, there are few attempts to consider the particle surface roughness in DEM \cite{nadimi2019effect}, not to mention this LSM. The similar works so far can be categorized into two groups: Refining the geometry, or improve the contact model \cite{pohlman2006surface,cavarretta2010influence,hare2013influence,wilson2017influence,zhao2018extended}. The former methodology is computationally expensive and restricted to meshing resolution \cite{hare2013influence,wilson2017influence}. For the latter, recently, researchers develop a micromechanical methodology for determining the effective interface stiffness, incorporating the surface roughness effects \cite{nadimi2019effect}. Specifically, Cavarretta et al. \cite{cavarretta2010influence}, Otsubo et al. \cite{otsubo2017influence}, and T. Zhao and Feng \cite{zhao2018extended} conduct the development of normal force-displacement relationships for rough surfaces based on the Hertz contact model. These models use the statistical approach of Greenwood and Tripp \cite{greenwood1967elastic} with a particle-scale roughness index. 

Hertz contact model provides a classic normal force-displacement relation for two smooth identical spheres in contact, given by
\begin{equation}
\label{eq1}
{F_N} = \frac{4}{3}{E^*}\sqrt {{R^*}} {u_{\rm{n}}}\sqrt {{u_{\rm{n}}}} ,
\end{equation}
where ${E^*}$ is the effective contact Young's modulus given by 
\begin{equation}
\label{eq2}
{E^*} = \frac{E}{{1 - {\nu ^2}}},
\end{equation}
and ${R^*}$ is the equivalent radius,
\begin{equation}
\label{eq3}
\frac{1}{{{R^*}}} = \frac{1}{{{R_1}}} + \frac{1}{{{R_2}}},
\end{equation}
in which $E$ is the elastic modulus and $\nu$ is Poisson's ratio of the contact surface.

\begin{figure}
\centering\includegraphics[width=0.6\linewidth]{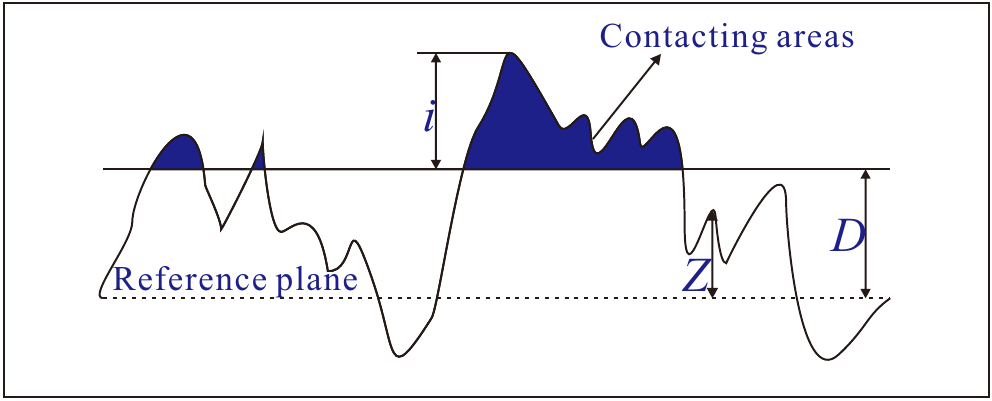}
\caption{Schematic of the contact of rough surfaces \cite{nadimi2019effect}}\label{f6}
\end{figure}

For a given normal displacement as shown in Fig. \ref{f6}, the normal load is lower for a rough particle than a smooth one, so an error function is added to Eq. (\ref{eq1}) as following \cite{nadimi2019effect},
\begin{equation}
\label{eq4}
{F_N} = \frac{4}{3}{E^*}\sqrt {{R^*}} {u_{\rm{n}}}\sqrt {{u_{\rm{n}}}} {\rm{ - }}\beta S_q^*{E^*}\sqrt {{R^*}S_q^*} {\rm{erf}}\left( {\alpha \frac{{{u_{\rm{n}}}}}{{S_q^*}}} \right),
\end{equation}
where $S_q^*$ is the effective value of two elements with the surface roughness of ${S_{{q_1}}}$ and ${S_{{q_2}}}$,
\begin{equation}
\label{eq5}
S_q^* = \sqrt {S_{{q_1}}^2 + S_{{q_2}}^2} ,
\end{equation}
with two extra roughness constants of $\alpha $ and $\beta $ from inputs.
\subsection{Smooth Joint Logic for Contacts Slip}
\label{s3.3}
\begin{figure}
\centering\includegraphics[width=1.0\linewidth]{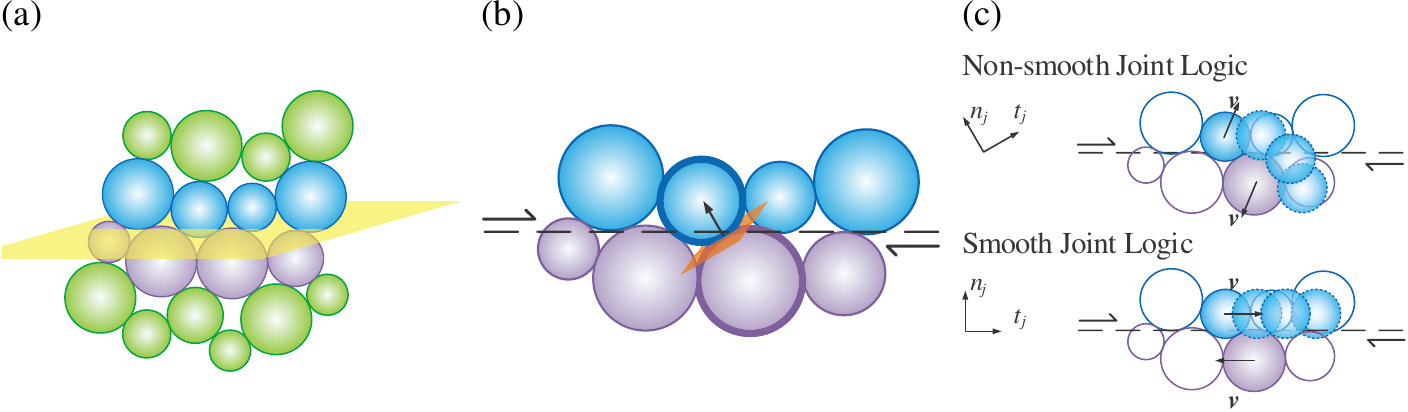}
\caption{(a) DEM-DFN model; (b) DEs crossing the discontinuity plane; (c) the movements without and with SJL of one DE \cite{liu2017mechanical}}\label{f7}
\end{figure}

Interfaces in discrete-based numerical models have been traditionally represented by debonding contacts along a line or plane \cite{liu2020stress} or assigning low strength and stiffness micro-properties to them \cite{ivars2011synthetic}. While, mesh sensitivity would cause artificial roughness and bumpiness arising from the particle-based material representation \cite{lisjak2014review}. This shortcoming is overcome by the development of the smooth-joint contact model (SJM) \cite{ivars2011synthetic}, which allows one to simulate a smooth interface regardless of the local particle topology \cite{lisjak2014review}. Based on this model, smooth joint logic (SJL) is introduced to simulate the pre-existing cracks and faults in DEM samples \cite{liu2017mechanical,potyondy2004bonded}. Figure \ref{f7} visualizes the SJL effect on the movement trajectories of DEs by shearing: Figure. \ref{f7}c shows the interaction across the pre-existing crack with classical and modified contact orientation, respectively.

\begin{figure}
\centering\includegraphics[width=1.0\linewidth]{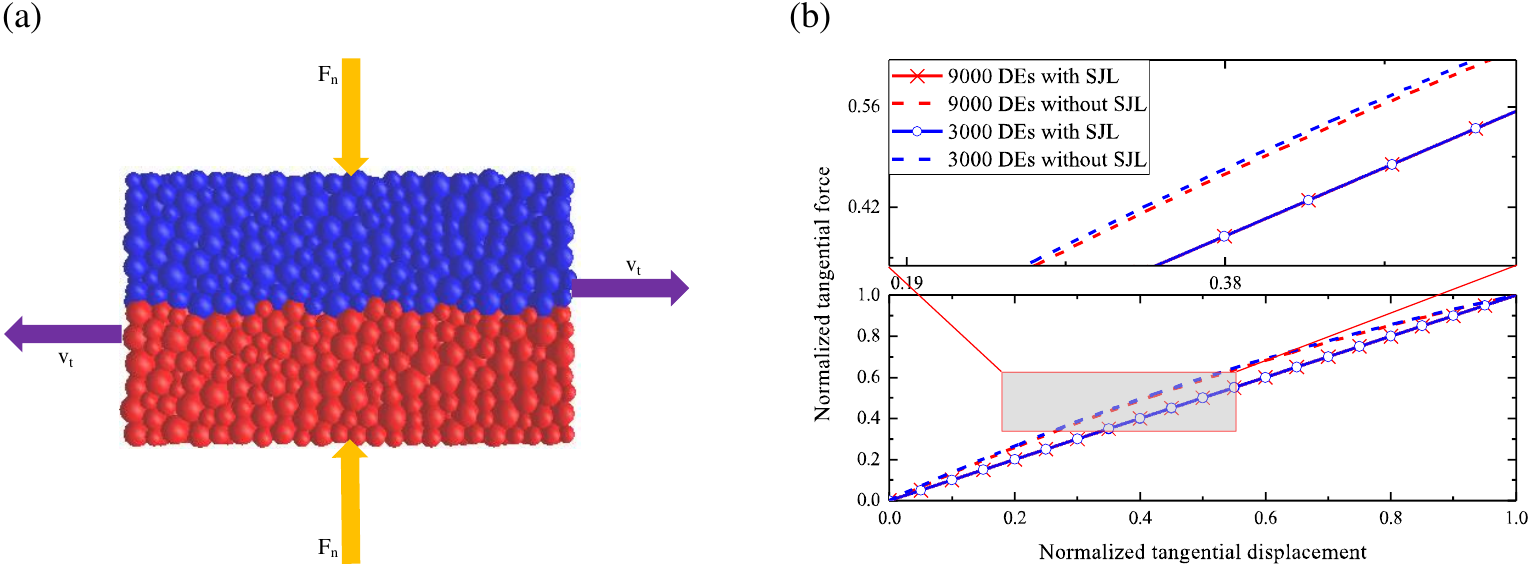}
\caption{Visualization of the SJL effect: (a) direct shear test with the orientations of the loads and movements; normalized tangential force-displacement variation}\label{f8}
\end{figure}

Figure \ref{f8} illustrates the SJL effect on the DEM simulation in direct shear tests. Specimens are subjected to a compression in vertical direction and then to shearing along the horizontal direction at a constant normal force (Fig. \ref{f8}a). We get the normalized tangential force-displacement responses: For the models with SJL, increasing elements do not affect the numerical results, whereas those without SJL do.

\subsection{LSM-DFN Coupling Model for Cracked Medium}
\label{s3.4}
\begin{figure}
\centering\includegraphics[width=1.0\linewidth]{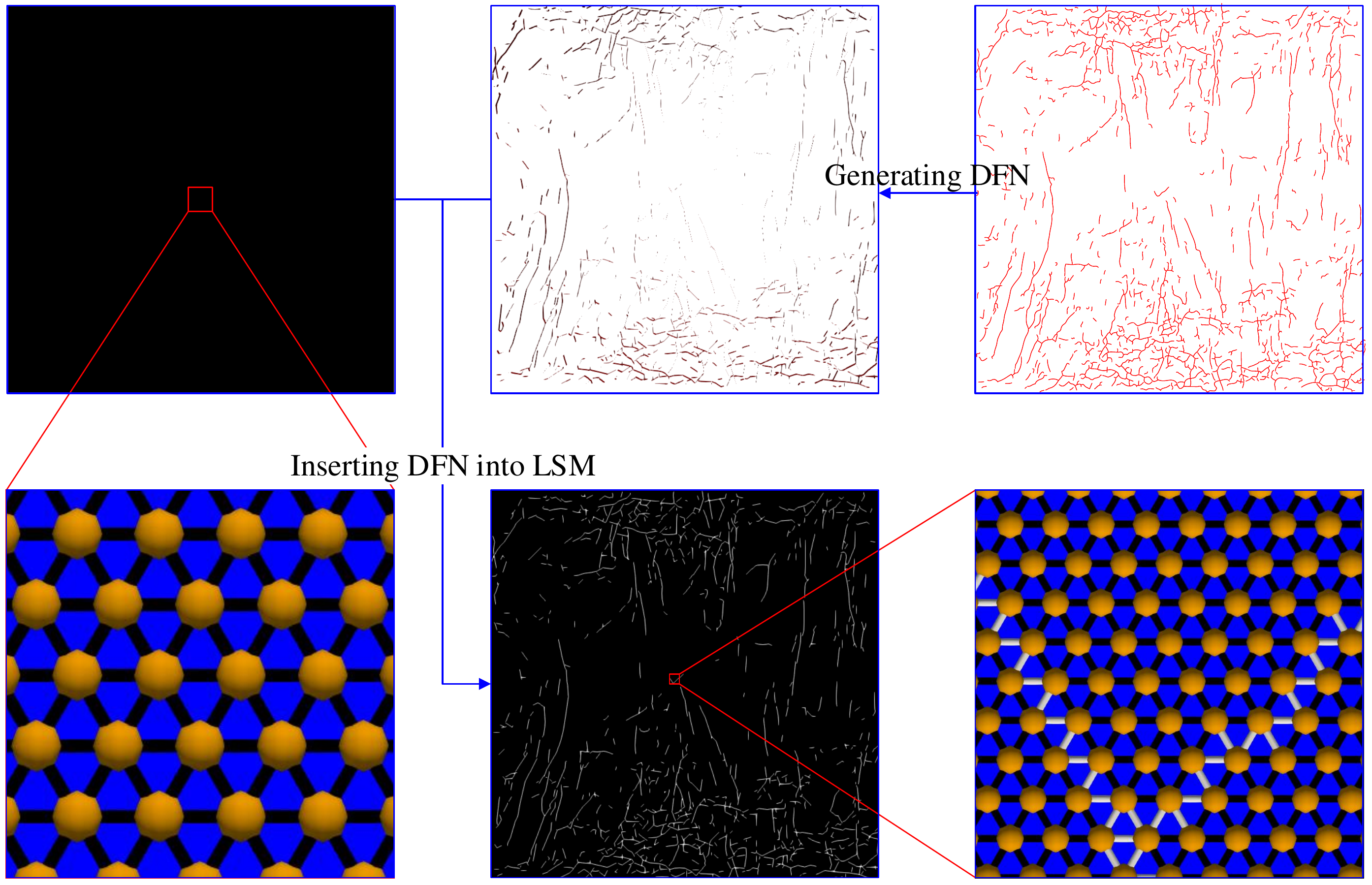}
\caption{LSM-DFN modeling for a naturally-fractured reservoir}\label{f9}
\end{figure}

DFN, is short for discrete fracture network referring to a computational model that explicitly represents the geometrical properties of each individual fracture, or the topological relationships between individual fractures and fracture sets \cite{lei2017use}. Increasingly, DFN modeling is being used in advanced engineering analyses to explicitly account for the presence of discontinuities in fractured rock masses \cite{alghalandis2017similarity,brady1993rock,jing2003review,jing2002numerical}. We propose an integrated LSM-DFN workflow for naturally-fractured reservoirs as shown in Fig. \ref{f9}. Firstly, we emulate the matrix as a linearly elastic, homogeneous, and isotropic material using the regular triangular modified LSM (more details given by N. Liu et al. \cite{liu2020modified,liu2017mechanical}). For a digital image of fractured rocks, the lattice nodes are set in an approximate zone represented by digital pixels. We extract automatically fracture attributes from the digital image based on the GrdHT algorithm to generate the DFNs. Then, the DFNs are inserted into the LSMs to form the LSM-DFN model for a naturally-fractured reservoir. The corresponding nodal interactions crossing the DNFs (white lines shown in Fig. \ref{f9}) are modified according to the research target as discussed in Sects. \ref{s3.2} and \ref{s3.3}.

\section{Validation and Calibration by Theoretical Prediction}
\label{s4}
The objective of this section is to validate the modified constitutive relations are introduced and implemented into the modified LSM scheme. By calibration process, LSM-DFN model could enable a high-accurate modeling by properly choosing reasonable meshing resolutions, damping values, and strain loading rates \cite{liu2020stress}. Moreover, by comparing with the theoretical predictions, the applicability of the effective medium theories could be further tested.
\subsection{Verification of Constitutive Laws}
\label{s4.1}
\begin{figure}
\centering\includegraphics[width=1.0\linewidth]{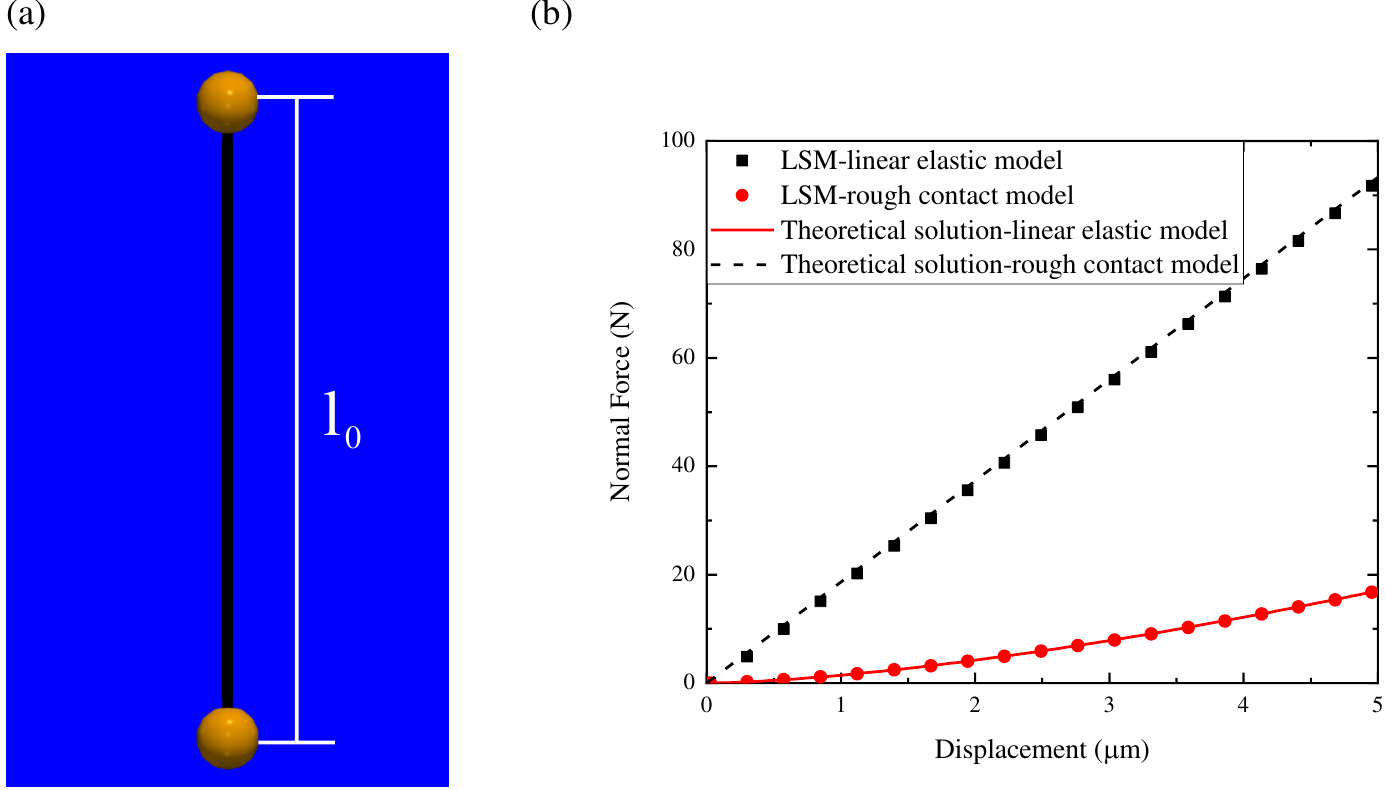}
\caption{Verification of LSM scheme: (a) Description of the LSM test; (b) Comparison between analytical and LSM solution}\label{f10}
\end{figure}

In order to verify the implement of the constitutive laws in this scheme, numerical tests are carried out according to the PFC2D manual \cite{cundall2004pfc2d} as shown in Fig. \ref{f10}. Two lattice nodes with two degrees of freedom are created, where one is fixed at the lower position and the other moves at a constant speed (Fig. \ref{f10}a) along the vertical direction. Assuming $E = 70{\kern 1pt} {\kern 1pt} {\rm{GPa}}$, $\nu  = 0.25$, $\alpha  = 0.01$, $\beta  = 10$, $R = 5 \times {10^{ - 4}}{\kern 1pt} {\kern 1pt} {\rm{m}}$, ${S_q} = 0.5{\kern 1pt} {\kern 1pt} {\rm{\mu m}}$, and ${K_{\rm{n}}} = {E^*}{R^*}$, a comparison of the contact force versus displacement for the analytical and LSM solution is shown in Fig. \ref{f10}b. From this figure, the LSM solution is shown to agree well with the theoretical prediction. This simple verification test guarantees the correctness of the calculation results for each contact between two lattice nodes, and thus increases confidence in the simulation results of the whole LSM scheme with a large number of nodes when using the rough contact model (Eq. (\ref{eq4})).

\subsection{LSM Modeling for Homogenous Media}
\label{s4.2}
\begin{figure}
\centering\includegraphics[width=0.7\linewidth]{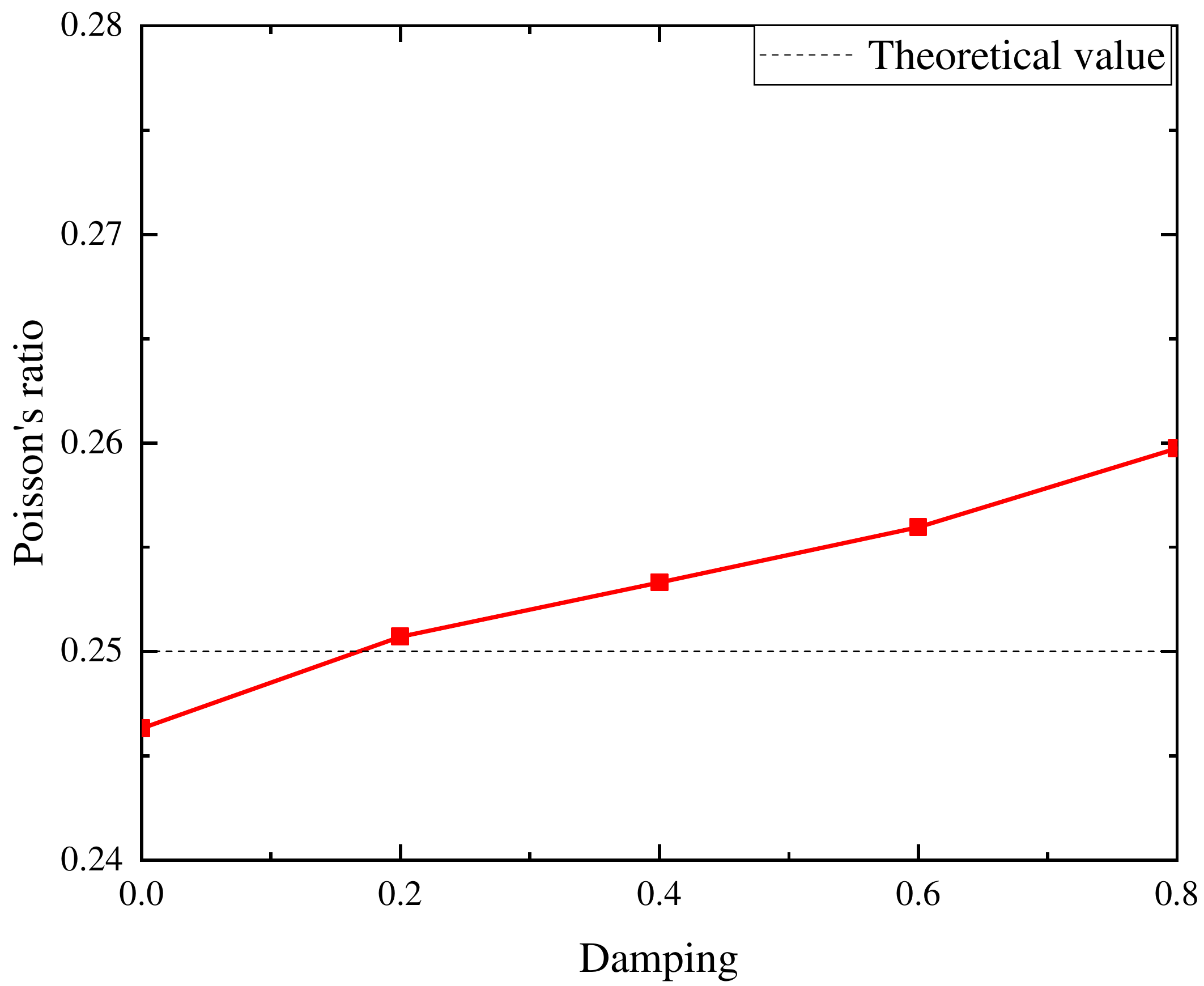}
\caption{Poisson’s ratio of the rock matrix changes with increasing numerical damping at the strain ratios of   ${{0.01} \mathord{\left/
 {\vphantom {{0.01} {\Delta t{\kern 1pt} {\kern 1pt} {{\rm{s}}^{ - 1}}}}} \right.
 \kern-\nulldelimiterspace} {\Delta t{\kern 1pt} {\kern 1pt} {{\rm{s}}^{ - 1}}}}$}\label{f11}
\end{figure}

Both numerical damping $\xi $ and strain rate $\dot \varepsilon $ may influence the numerical results as mentioned in our previous work \cite{liu2020stress}. To enable a high-accurate LSM modeling, we perform uniaxial compression tests with a reasonable meshing resolution of ${D \mathord{\left/ {\vphantom {D d}} \right. \kern-\nulldelimiterspace} d} = 400$ and different set-up parameters. The relation between the Poisson's ratios of the macroscopic material properties and lattice node parameters is expressed as (details in Appendix \ref{a2}),
\begin{equation}
\label{eq6}
\nu  = \frac{{{K_{\rm{n}}} - {K_{\rm{s}}}}}{{3{K_{\rm{n}}} + {K_{\rm{s}}}}} = \frac{{1 - \xi }}{{3 + \xi }},
\end{equation}
where $\xi  = \frac{{{K_{\rm{s}}}}}{{{K_{\rm{n}}}}}$ is the ratio of the shear to normal stiffness. Figure \ref{f11} exhibits the obtained Poisson's ratios of the same LSM sample at an 'optimal' strain rates of ${{0.01} \mathord{\left/
 {\vphantom {{0.01} {\Delta t{\kern 1pt} {\kern 1pt} {{\rm{s}}^{ - 1}}}}} \right.
 \kern-\nulldelimiterspace} {\Delta t{\kern 1pt} {\kern 1pt} {{\rm{s}}^{ - 1}}}}$  
\cite{liu2020stress}. With increasing numerical damping $\xi $, the Poisson's ratio $\nu$ becomes larger for this given material. We could see the minimum difference happens at the conditions that numerical damping equals 0.2.

\subsection{LSM-DFN Models with Uniformly Oriented Cracks}
\label{s4.3}
\begin{figure}
\centering\includegraphics[width=1.0\linewidth]{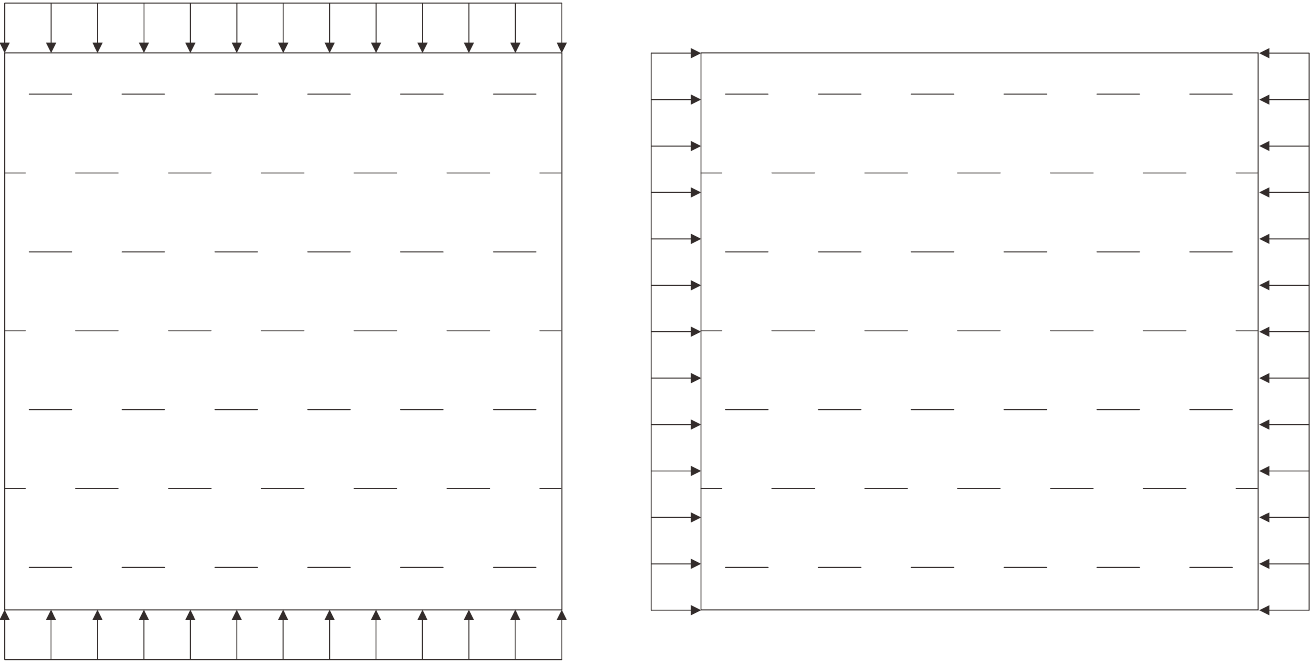}
\caption{Loading patterns along $y$ (left) and $x$ (right) axes}\label{f12}
\end{figure}

It is common in rocks that the cracks have preferred orientations, which might reflect the stress history of the rocks due to stress-induced anisotropy  \cite{liu2020stress}. Generally, such situation makes rocks heterogeneous and anisotropic. Some frequently used theoretical models for the effective elastic stiffness of cracked media are reviewed in Appendix \ref{a3}. To verify the numerical algorithm and enrich the knowledge of the limitations for those effective medium theories, we emulate different LSM-DFN models with various crack lengths and crack numbers by uniaxial compression tests. Specifically, isotropic rock matrix with horizontally aligned cracks, making the fractured system transversely isotropic with a vertical symmetry axis (VTI) \cite{zhao2015characterizing}. Figure \ref{f12} shows the two kinds of loading patterns along the $y$ and $x$ directions for the LSM-DFN specimen, respectively. Given that the numerical models are in two dimensions, the constitutive relations can be given by,
\begin{equation}
\label{eq7}
\left( {\begin{array}{*{20}{c}}
{{\varepsilon _{11}}}\\
{{\varepsilon _{22}}}\\
{{\varepsilon _{12}}}
\end{array}} \right) = \left( {\begin{array}{*{20}{c}}
{\frac{1}{{{E_{11}}}}}&{ - \frac{{{\nu _{21}}}}{{{E_{22}}}}}&0\\
{ - \frac{{{\nu _{12}}}}{{{E_{11}}}}}&{\frac{1}{{{E_{22}}}}}&0\\
0&0&{\frac{1}{{2{G_{12}}}}}
\end{array}} \right)\left( \begin{array}{l}
{\sigma _{11}}\\
{\sigma _{22}}\\
{\sigma _{12}}
\end{array} \right)
\end{equation}
where the Young's moduli and the pair of Poisson's ratios are defined by,
\begin{equation}
\label{eq8}
{E_{11}} = \frac{{{\sigma _{11}}}}{{{\varepsilon _{11}}}},
\end{equation}
\begin{equation}
\label{eq9}
{E_{22}} = \frac{{{\sigma _{22}}}}{{{\varepsilon _{22}}}};
\end{equation}
\begin{equation}
\label{eq10}
{\nu _{{\rm{12}}}} =  - \frac{{{\varepsilon _{22}}}}{{{\varepsilon _{11}}}},
\end{equation}
\begin{equation}
\label{eq11}
{\nu _{{\rm{21}}}} =  - \frac{{{\varepsilon _{11}}}}{{{\varepsilon _{22}}}};
\end{equation}
and
\begin{equation}
\label{eq12}
{E_{11}}{\nu _{21}} = {E_{22}}{\nu _{12}},
\end{equation}
owing to the known symmetry condition. The Young's moduli (${E_{11}}$ and $E_{22}$) and the Poisson's ratios ( $\nu _{12}$ and ${\nu _{21}}$) could be calculated from the slopes of the curves for compressive stress and transverse strain to the corresponding axial strain. Giordano and Colombo \cite{giordano2007effects} used a unified theory covering all the orientation distributions between the random and the parallel ones. For this case, cracks are aligned with a horizontal direction, the effective stiffness are,
\begin{equation}
\label{eq13}
{C_{1111}} = \frac{{(1 - \nu )(1 + 2\varepsilon )E}}{{D(1 + \nu )}},
\end{equation}
\begin{equation}
\label{eq14}
{C_{2222}} = \frac{{(1 - \nu )E}}{{D(1 + \nu )}};
\end{equation}
\begin{equation}
\label{eq15}
{C_{{\rm{1122}}}}{\rm{ = }}{C_{{\rm{2211}}}}{\rm{ = }}\frac{{\nu E}}{D},
\end{equation}
where
\begin{equation}
\label{eq16}
D = \left( {2{{(1 - \nu )}^2}\varepsilon  + 1 - 2\nu } \right)\left( {1 + \nu } \right),
\end{equation}
and crack density $\varepsilon $ is defined by
\begin{equation}
\label{eq17}
\varepsilon  = \frac{{N{b^2}}}{A},
\end{equation}
in which $N$, $b$, and $A$ are the crack number, the half-length of a crack, and the area of a specimen, respectively. 

\begin{figure}
\centering\includegraphics[width=0.7\linewidth]{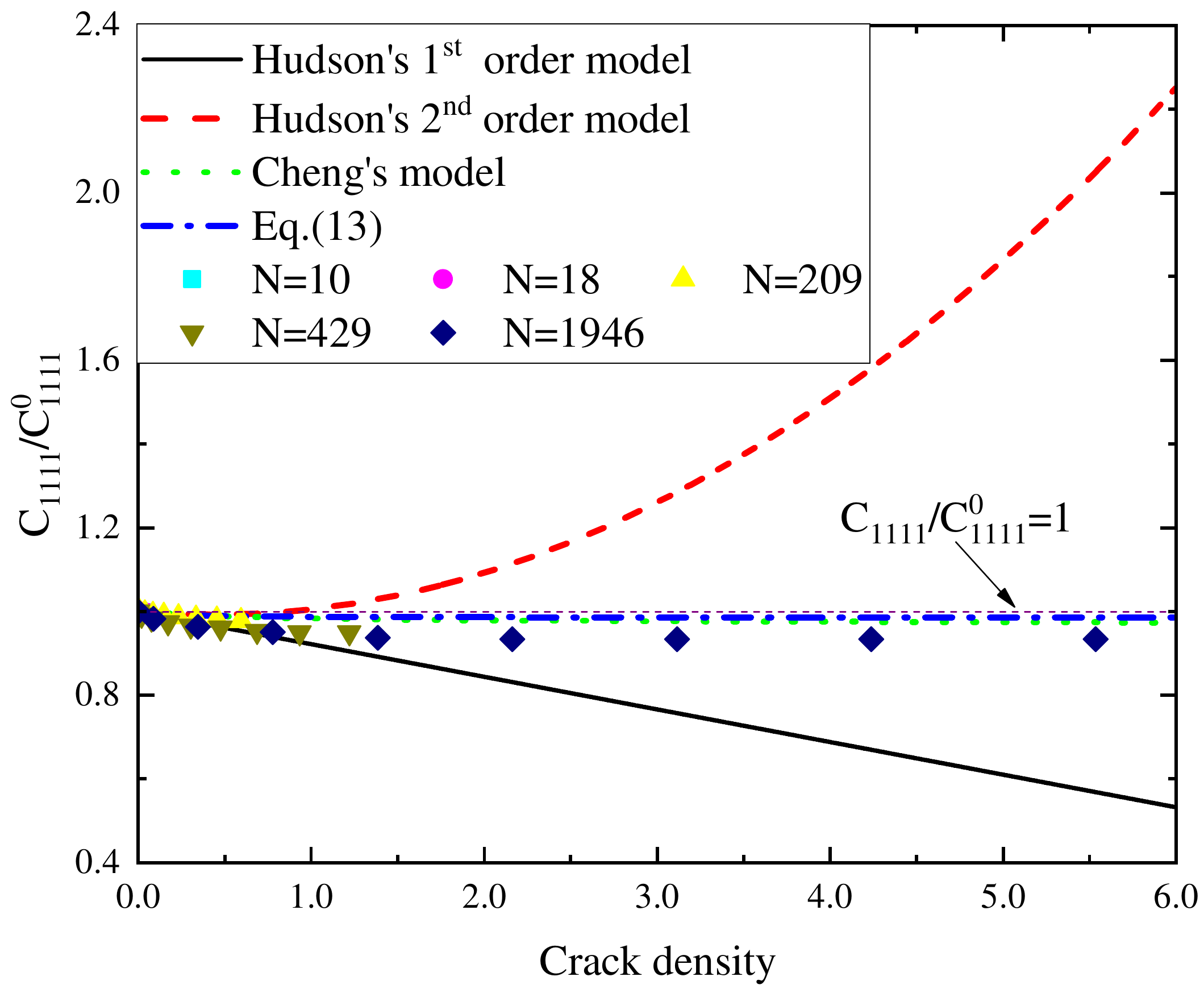}
\caption{Comparisons of normalized effective stiffness $\frac{{{C_{1111}}}}{{C_{1111}^0}}$ with respect to crack density predicted by different effective medium theories}\label{f13}
\end{figure}

Figure \ref{f13} displays a normalized effective stiffness of $\frac{{{C_{1111}}}}{{C_{1111}^0}}$ for the numerical results calculated by LSM-DFN models with the crack numbers of 10, 18, 209, 429 and 1946, and comparisons of those values by the theoretical predictions of Hudson's first- and second- order models, Cheng's models and the formulae proposed by Giordano and Colombo \cite{giordano2007effects}. As expected, all the values do agree with each other within a dilute limit. While, with a higher crack density, there appear significant differences: The theoretical values of first- and the second-order expansions are divergent, whilst the expressions of Cheng's model and Eq. (\ref{eq13}) fall between the curves of the first- and second-order Hudson's models, and the normalized values calculated by  Eq. (\ref{eq13}) are larger than those of Cheng's model but less than 1.0 with increasing crack density.

\begin{figure}
\centering\includegraphics[width=0.7\linewidth]{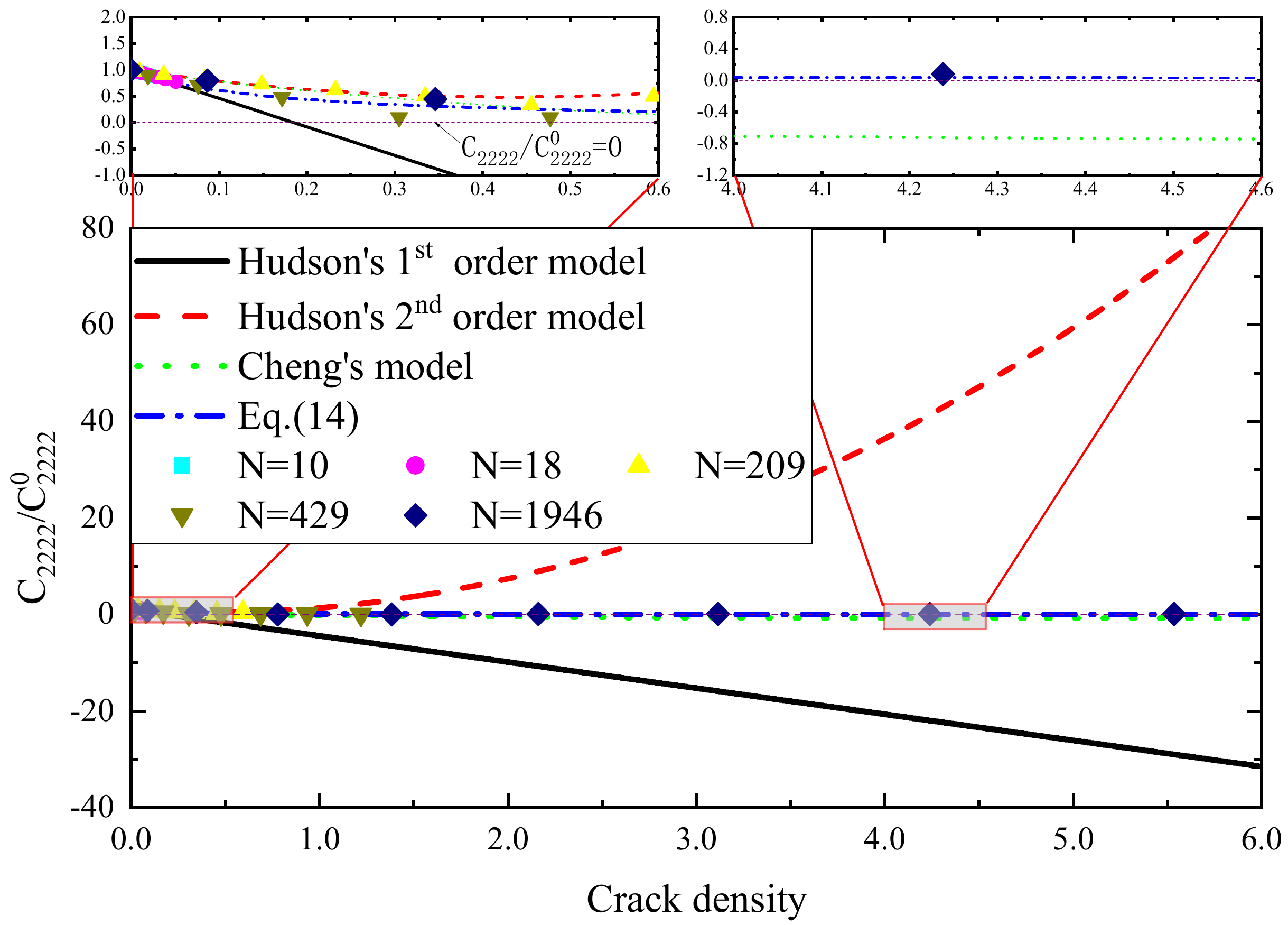}
\caption{Comparisons of normalized effective stiffness $\frac{{{C_{2222}}}}{{C_{2222}^0}}$ with respect to crack density predicted by different effective medium theories}\label{f14}
\end{figure}

\begin{figure}
\centering\includegraphics[width=0.7\linewidth]{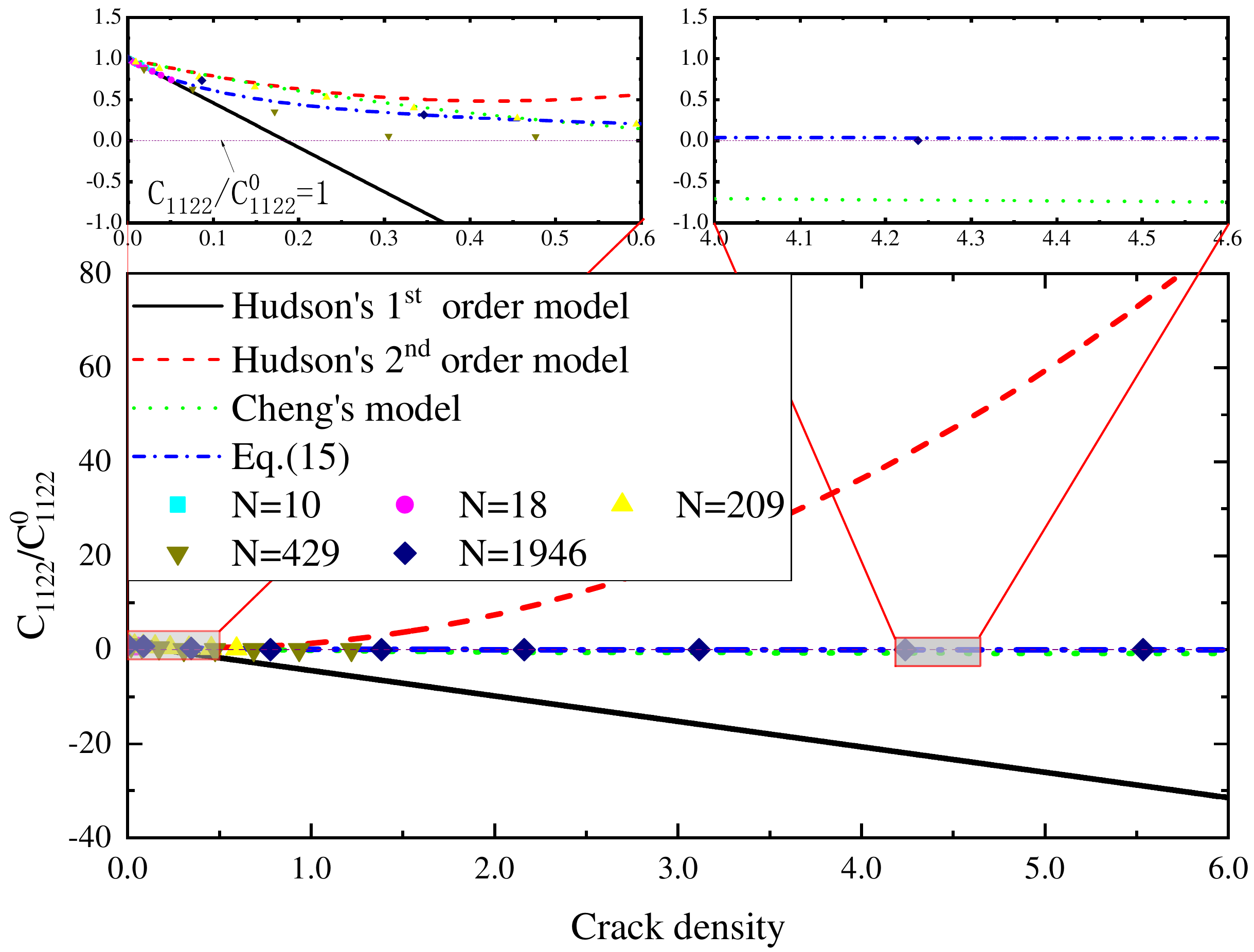}
\caption{Comparisons of normalized effective stiffness $\frac{{{C_{1122}}}}{{C_{1122}^0}}$ with respect to crack density predicted by different effective medium theories}\label{f15}
\end{figure}

Figures. \ref{f14} and \ref{f15} compare the numerical results with those theoretical predictions of the effective stiffness coefficients $\frac{{{C_{2222}}}}{{C_{2222}^0}}$ and $\frac{{{C_{1122}}}}{{C_{1122}^0}}$, respectively. Along with the tendency of $\frac{{{C_{1111}}}}{{C_{1111}^0}}$, within a low crack density nearly less than 0.1, the four approximate solutions almost coincide and are close to those of LSM-DFN modeling. With a higher crack density, the first-order Hudson's solutions significantly deviate from them; the values from the second-order Hudson's models allow to extend the validity of the theory to a higher crack density up to 1.0. However, when the crack density is larger than 1.0, the curve of first-order Hudson's expression monotonically drops below zero, indicating that a cracked solid has negative elastic stiffness, which is unreasonable. Meanwhile, the tendency of the second-order Hudson's theory to produce abnormally high effective elastic constants stiffer than the intact one. By Cheng's model, this elastic stiffness converges to a certain value but still less than zero, which is against the physical fact. Our numerical results agree well with the predictions by Eqs. (\ref{eq14}) and (\ref{eq15}), tending to be zero with increasing crack density. For this case study, the expressions developed by Giordano and Colombo  \cite{giordano2007effects} may be superior to the other theoretical models.

\begin{figure}
\centering\includegraphics[width=1.0\linewidth]{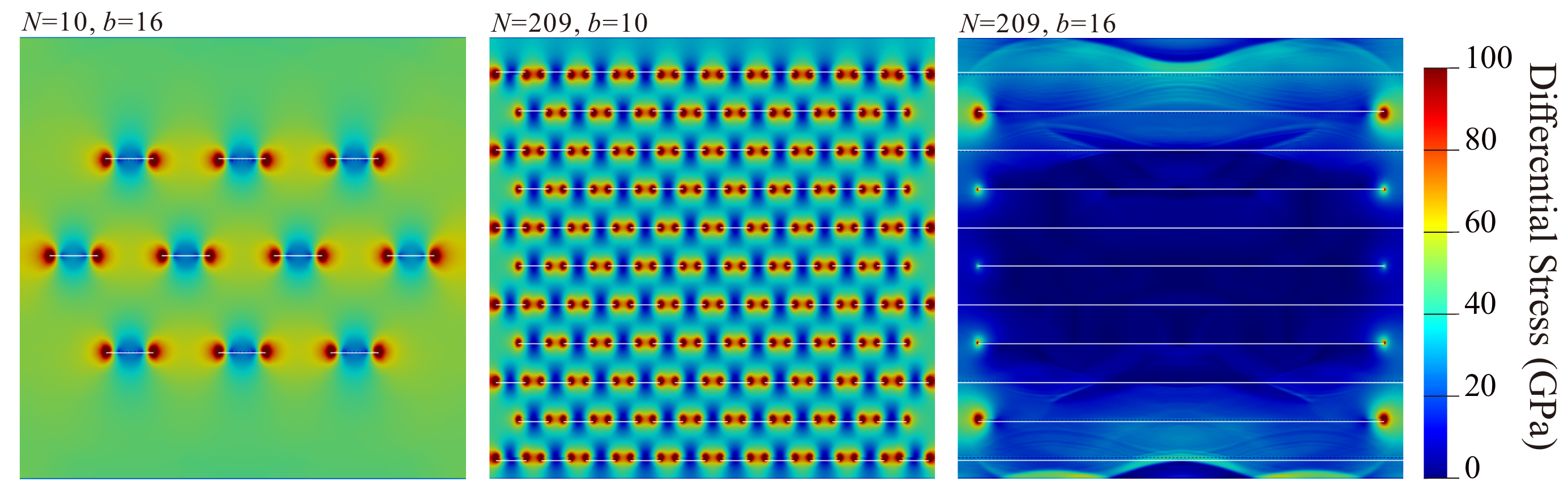}
\caption{Differential stress fields at ${\varepsilon _{yy}}$ = 2.5\% for LSM-DFN models with different crack numbers and crack lengths}\label{f16}
\end{figure}

As expected, the cracks orthogonal to the compressive axis make more significant effect on the elastic stiffness coefficients $\frac{{{C_{2222}}}}{{C_{2222}^0}}$ and $\frac{{{C_{1122}}}}{{C_{1122}^0}}$ than that of $\frac{{{C_{1111}}}}{{C_{1111}^0}}$ which is related to the Young's modulus along $x$ direction. For this reason, Fig. \ref{f16} shows the differential stress fields of some selected LSM-DFN models under the compressive strain of 2.5\% along $y$ axis. From this figure, local stress distributions around the crack tips could be visualized. Two types of the stress disturbances caused by cracks can be caught, including the stress concentration, or known as stress amplification \cite{grechka2006effective,zhao2015characterizing} and stress shielding occurring around the crack faces for a low crack density, such as the specimen with the crack number of 10, and the crack half-length of 16 shown in Fig. \ref{f16}. Within a high-crack-density model, the local stress fields are disturbed by dense crack distribution, and the differential stresses of the crack faces might not be free. The phenomenon may further explain the reason why the application of both first- and second-order Hudson's expressions has a dilute limit. Obviously, this numerical approach shows advantages over the theoretical solutions for predicting the mechanical properties of cracked solids.

\section{Numerical Results and Discussions}
\label{s5}
In this section, we hold a numerical simulation up as an example, showing the proposed workflow of LSM-DFN modeling for a rocky outcrop, containing multiple cracks incorporating rough contact deformation. The dependence of fracture networks and micro-scale surface roughness on the elastic characteristics is quantitatively discussed here by the simulation results of uniaxial compression tests along $y$ and $x$ axes.
\subsection{Models and Loading Paths}
\label{s5.1}
Distinct from isotropic media, only by two elastic constants, such as Young's moduli and Poisson's ratios, it's insufficient to characterize the elastic properties of an anisotropic rock. Therefore, we perform a comprehensive series of numerical simulations to explore the elastic constants related to compression tests. A two-dimensional square domain with a size of $600 \times 600{\kern 1pt} {\kern 1pt} {\kern 1pt} {{\rm{m}}^2}$ is used and discretized into mass nodes connecting with normal and shear springs. In addition, the fracture patterns from an outcrop photograph are represent by DFNs. We could estimate the crack density of the model by Eq. (\ref{eq17}), namely 0.959126. The physical and mechanical properties of lattice nodes are reported in Table \ref{t1}. Each simulation lasts for approximately 3 hours for 183770 lattice nodes using a single-core CPU with 3.7 GHz.

\begin{table}
\centering
\caption{Material properties in the simulation}\label{t1}
\begin{tabular}{l l}
\hline
Parameter &Value\\
\hline
Density [$\rho $ (kg/m$^3$)]&2500\\
Young's modulus [$E$ (GPa)]&70\\
Poisson's ratio [$\nu$ (-)]& 0.25\\
Roughness constants [$\alpha$ and $\beta$ (-)]&0.1, 10\\
Surface roughness [${S_q}$ (mm)] & 0.1\\
\hline
\end{tabular}
\end{table}

\begin{figure}
\centering\includegraphics[width=1.0\linewidth]{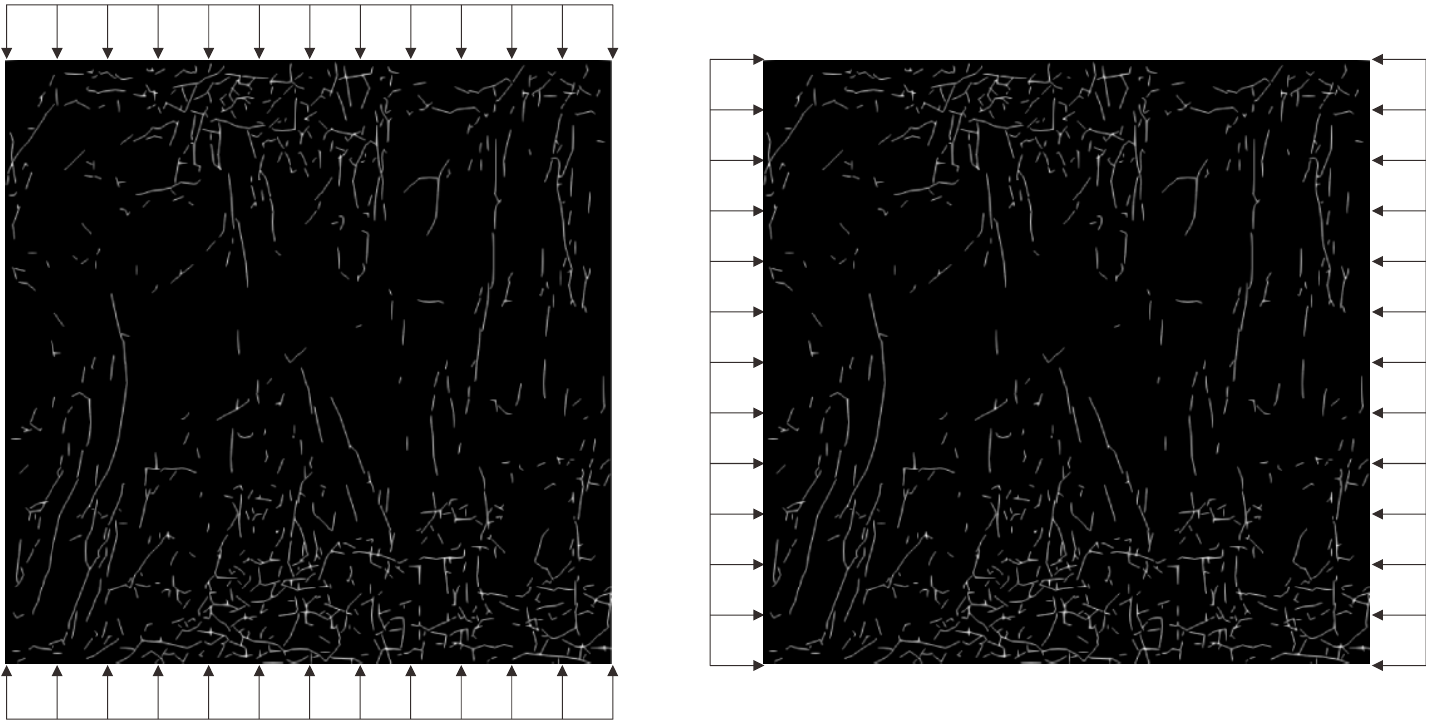}
\caption{Loading patterns of LSM-DFN model: along $y$ axis (Model 1, left); along $x$ direction (Model 2, right)}\label{f17}
\end{figure}

As shown in Fig. \ref{f17}, the resulting LSM-DFN model is compressed along two directions, respectively. It is worth mentioning that ${E_{yy}}$ and ${\nu _{yx}}$ denote the Young's modulus and Poisson's ratio from the compression test along $y$ direction denoted as Model 1; on the other hand, ${E_{xx}}$ and ${\nu _{xy}}$
are calculated by the test along $x$ axis as Model 2.

\subsection{Effects of Fracture Attributes and Micro-surface Roughness}
\label{s5.2}

\subsubsection{Effect on Stress-Strain Response}
\label{s5.2.1}
\begin{figure}
\centering\includegraphics[width=0.7\linewidth]{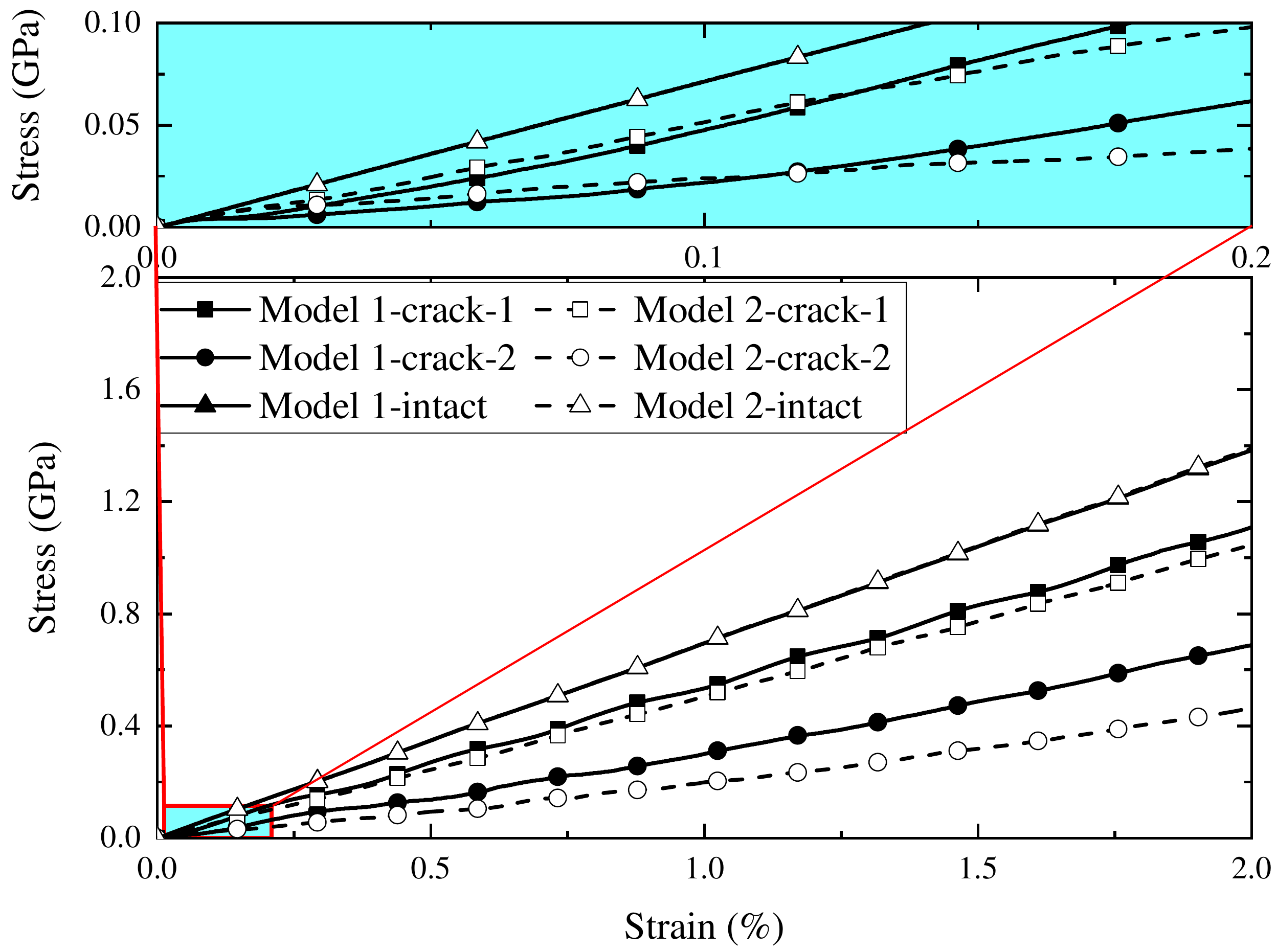}
\caption{Stress responses for cracked and intact media under two loading paths}\label{f18}
\end{figure}

Figure \ref{f18} show the stress-strain responses, in which Model 1-crack-1, Model 1-crack-2, and Model 1-intact mean LSM-DFN modeling for the cracked solid with modified contact relations (Eq. (\ref{eq4})) and debonding contacts used by N. Liu and Fu \cite{liu2020stress} for the fracture surfaces, and LSM modeling for the homogeneous medium compressed along $y$ axis, respectively; Model 2-crack-1, Model 2-crack-2, and Model 2-intact denote those models loaded along $x$ axis, correspondingly. As shown in this figure, the response curves of crack-free models, Model 1-intact and Model 2-intact, compressed in two directions are coincident, which indicates that the triangular lattice arrangement is suitable for emulating isotropic media, and the implement of the algorithm could be further verified. For the cracked models, the slopes of stress-strain curves are lower than the value of the corresponding intact specimen as theories predicted. At the initial stage, the curves of Model 2-crack are steeper than those of Model 1-crack. While, as the axial strain increases, the axial-stress response curves along the $y$-axis is higher than those along the $x$-axis. This tendency could be attributed to the unevenly-distributed cracks and fracture networks: More cracks occur near the loading position along the horizontal direction and there is a relatively small crack density in the central part of the sample as illustrated in Figs. \ref{f3} and \ref{f4}. The slopes of the curves of Model 1-crack-1 and Model 2-crack-1 are significantly higher than those of Model 1-crack-2 and Model 1-crack-2, respectively, which shows that traditional cracked models with debonding contacts usually underestimate the elastic stiffness of the cracked media. It further illustrates the necessity of introducing the normal force-displacement relation for rough fracture surface.

\subsubsection{Effect on the Effective Elastic Stiffness and Poisson's Ratios}
\label{s5.2.2}
Table \ref{t2} lists the normalized values of the effective Young's moduli and Poisson's ratios, where those of intact models are treated as the references. Here, we use a parameter defined by Thomsen \cite{thomsen1986weak} to qualify the anisotropic elasticity as follows,
\begin{equation}
\label{eq18}
\gamma  = \frac{{{C_{1111}} - {C_{2222}}}}{{2{C_{2222}}}}.
\end{equation}
According to the formula, we could get the the anisotropy parameters of LSM-DFN models with rough contact deformation and debonding contacts, 0.0331 and 0.2592, respectively. It can be seen that, due to the introduction of micro-scale surface roughness, the degree of anisotropy of the fractured medium is reduced.
\begin{table}
\centering
\caption{Effective Young’s moduli and Poisson's ratios}\label{t2}
\begin{tabular}{l l l}
\hline
Model ID & ${E \mathord{\left/
 {\vphantom {E {{E_{{\rm{intact}}}}}}} \right.
 \kern-\nulldelimiterspace} {{E_{{\rm{intact}}}}}}$ (\%) & ${\nu  \mathord{\left/
 {\vphantom {\nu  {{\nu_{{\rm{intact}}}}}}} \right.
 \kern-\nulldelimiterspace} {{\nu_{{\rm{intact}}}}}}$ (\%)\\
\hline
Model 1-crack-1&79.16&99.28\\
Model 1-crack-2&46.66&26.87\\
Model 2-crack-1& 74.24&92.49\\
Model 2-crack-2&30.73&29.12\\
\hline
\end{tabular}
\end{table}
\subsubsection{Effect on Coordination Number Variation}
\label{s5.2.3}
For discrete-based numerical methods, the average number of contacting particles is named as coordination number. It is well known that stiffness is related to the coordination number \cite{rasp2013discrete}. Figure 19 presents the evolution of the average coordination number with respect to mean stress. Coordination number for the intact LSM sample is about 5.9813, and for the LSM-DFN models with debonding contacts and with rough contact deformation remain in the range 5.981 to 6.020, and 5.981 to 5.978, respectively, as mean pressures increase from 0 MPa to 1000 MPa. There are slight decreases in the cases of Model 1-crack-1 and Model 2-crack-1. It might be inferred that the introduction of the normal force–displacement relation for rough fracture surface hinders the closure of cracks during the compression tests, and meanwhile, promotes crack opening in the transverse direction. The slope of Model 2-crack-2 is much steeper than that of Model 1-crack-2, which hints that the apertures in $x$ direction are higher than those along the $y$ axis.

\begin{figure}
\centering\includegraphics[width=0.7\linewidth]{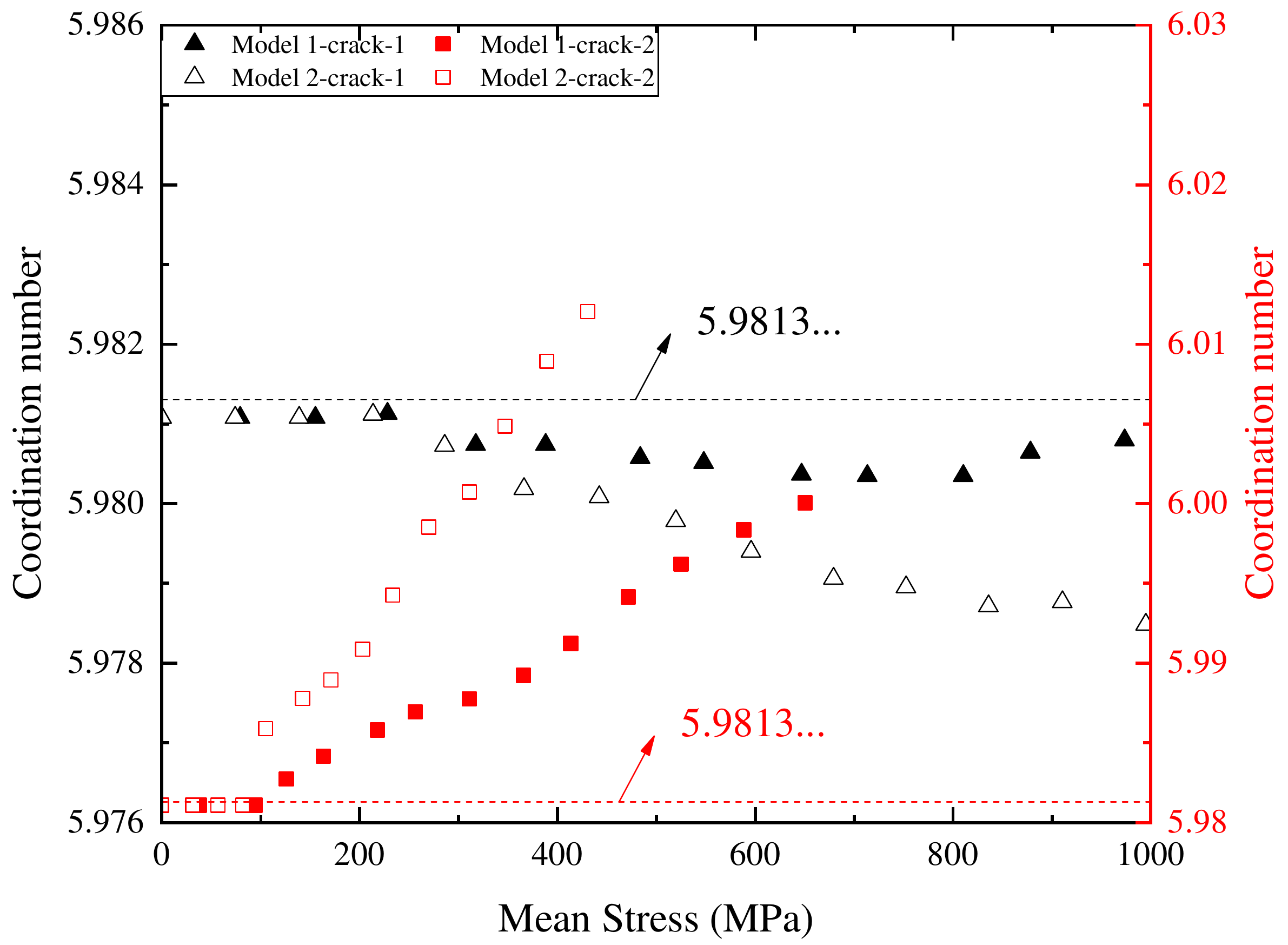}
\caption{Average coordination numbers at different stress levels}\label{f19}
\end{figure}

\subsubsection{Effect on Stress States}
\label{s5.2.4}
\begin{figure}
\centering\includegraphics[width=1.0\linewidth]{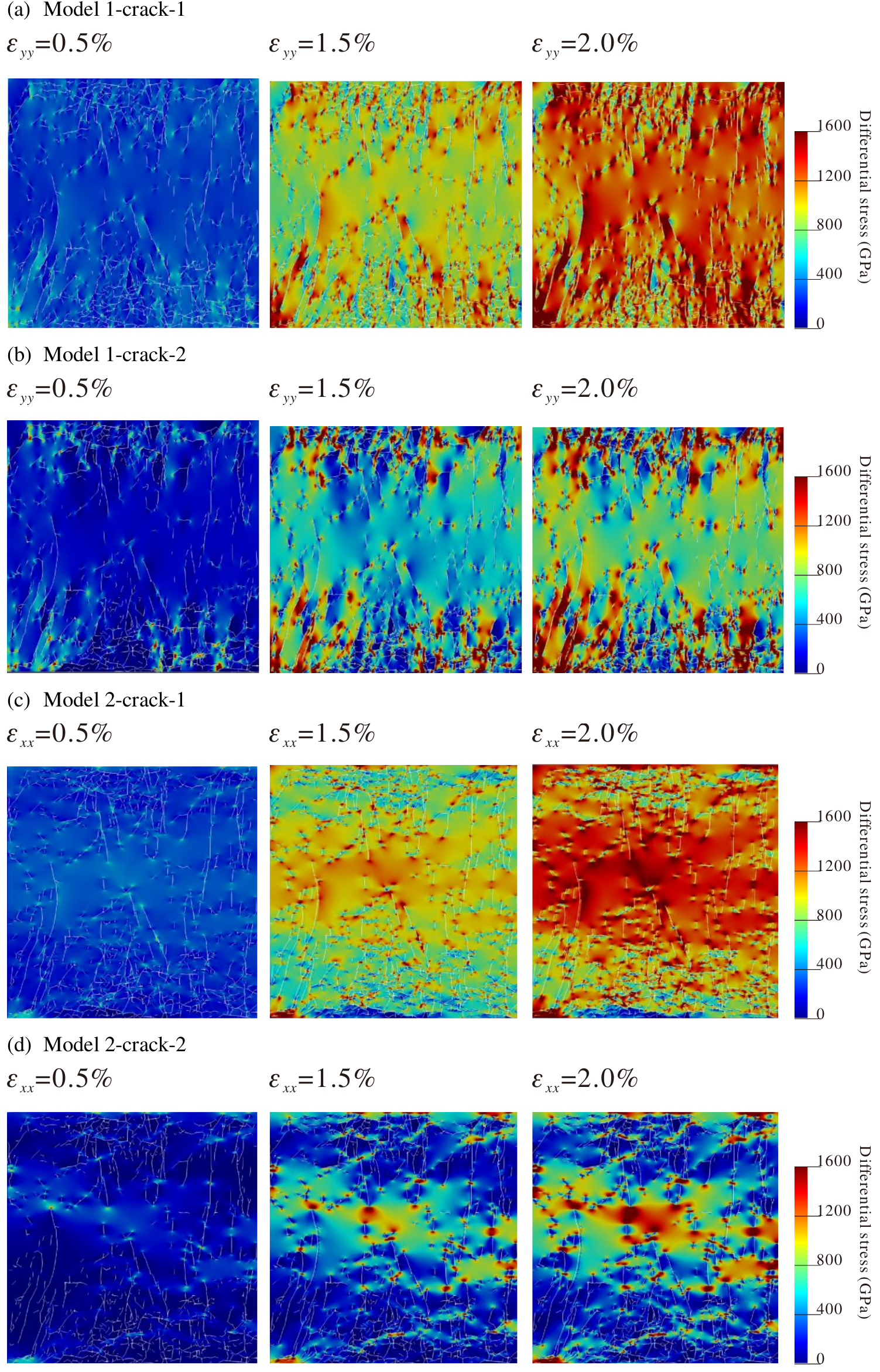}
\caption{Differential stress fields at different strain states for LSM-DFN models of the realistic cracked medium}\label{f20}
\end{figure}

Stress state of the medium can be enhanced or reduced by the surrounding fractures, depending on the mode of deformation, location, and fracture orientation \cite{thomas2017quantification}. As a result, a thorough understanding of the stress disturbances is essential for predicting the mechanical behaviors of a fractured system. Differential stress is usually used to assess whether tensile or shear failure will occur decided by the introduced failure criterion \cite{liu2020stress}. Figure. \ref{f20} displays the differential stress fields at different strain states for LSM-DFN models of the realistic cracked medium. From the figure, we can see the differential stress states under two load paths are different, especially for the models with debonding contacts. The stress concentration zones with local stresses greater than the background are generated at tips of these cracks orthogonal to the compressive axis, whilst local minimum stress tend to appear around the fracture surfaces \cite{liu2020stress}. Figure. \ref{f20}b and \ref{f20}d display more drastic changes in the local stress field disturbed by numerous cracks: Stress amplification around cracks dominates the stress states. From Fig. \ref{f20}a and \ref{f20}c, we could see that the whole differential stress fields show higher stress values, which may imply the introduction of the modified contact relation for the fracture surfaces stiffens the medium and weakens the stress concentration phenomenon of the fractured medium.

\section{Conclusions}
\label{s6}
In this study, an integrated numerical scheme accounting rough contact deformation is presented by coupling modified LSM and DFN modeling for predicting effective mechanical properties of a realistic outcrop. Those extremely-complex fracture networks are automatically extracted based on GrdHT algorithm. In this study, the effect of surface roughness on contact interaction is quantified and implemented in the modified LSM. The reason why the SJL is needed in the discrete-based model capable of surface roughness, which seems paradoxical, is explained systematically. The developed framework is employed to quantitatively investigate the effect of fracture attributes and micro-scale surface roughness on compression mechanism. Accordingly, the major conclusions and merits of this scheme are organized as follows.

1. This proposed LSM-DFN workflow is verified to be a surrogate choice to numerically investigate the cracked media in an effective way by comparing with the theoretical predictions.

2. The applicability of the effective medium theories could be tested. For this two-dimensional issue (every node with only two degrees of freedom), four mentioned theoretical predictions almost coincide with each other when the crack density within 0.1; with a higher crack density, the expression proposed by Giordano and Colombo \cite{giordano2007effects} may be a better candidate for estimating the effective elastic characteristics of cracked solids.

3. Traditional cracked models with debonding contacts usually underestimate the elastic stiffness of the cracked media. Rough contact deformation caused by micro-scale surface roughness tends to reduce the degree of anisotropy and to weaken the stress concentration phenomenon. Moreover, it may hinder the closure of cracks with the normal along the compressive axis, and promote crack opening in the transverse direction.

\appendix
\renewcommand{\thesection}{Appendix \Alph{section}}
\section{Material Properties and Lattice Node Parameters}
\setcounter{table}{0}   
\setcounter{figure}{0}
\setcounter{equation}{0}
\renewcommand{\thetable}{A\arabic{table}}
\renewcommand{\thefigure}{A\arabic{figure}}
\renewcommand{\theequation}{A\arabic{equation}}
\label{a2}
In two dimensions, the basic idea in setting up the spring network models is based on the equivalence of strain energy stored in a unit cell (Fig. \ref{fA-1}) with the area ${A_{{\rm{cell}}}}$ of a network \cite{22}
\begin{equation}
\label{eq34}
{U_{{\rm{cell}}}} = {U_{{\rm{continuum}}}},
\end{equation}
where the energies of the cell and its continuum equivalent, respectively, are 
\begin{equation}
\label{eq35}
{U_{{\rm{cell}}}} = \sum\limits_{n = 1}^6 {\frac{1}{2}\left( {{K_{\rm{n}}}{{\left( {u_{\rm{n}}^i} \right)}^2} + {K_{\rm{s}}}{{\left( {u_{\rm{s}}^i} \right)}^2}} \right)} ,
\end{equation}
\begin{equation}
\label{eq36}
{U_{{\rm{continuum}}}} = \frac{1}{2}{\bm{\varepsilon }} \cdot {\bf{C}} \cdot {\bm{\varepsilon }},
\end{equation}
in which superscript $i$ in Eq. (\ref{eq35}) stands for the ${i^{{\rm{th}}}}$ interaction. Then, the fourth-order effective stiffness tensor ${\bf{C}}$ can be derived by
\begin{equation}
\label{eq37}
{C_{ijkl}} = \frac{{{\partial ^2}{\omega _{{\rm{cell}}}}}}{{\partial {\varepsilon _{ij}}\partial {\varepsilon _{kl}}}}.
\end{equation}
Therefore, we obtain the expressions for the two Lam\'e parameters $\lambda $ and $\mu $, 
\begin{equation}
\label{eq38}
\lambda {\rm{ = }}\frac{{\sqrt {\rm{3}} }}{4}\left( {{K_{\rm{n}}} - {K_{\rm{s}}}} \right).
\end{equation}
\begin{equation}
\label{eq39}
\mu {\rm{ = }}\frac{{\sqrt {\rm{3}} }}{4}\left( {{K_{\rm{n}}} + {K_{\rm{s}}}} \right).
\end{equation}
And for a two-dimensional (2D) isotropic elastic medium, we also have
\begin{equation}
\label{eq40}
\lambda  = \frac{{E\nu }}{{1 - {v^2}}},
\end{equation}
\begin{equation}
\label{eq41}
\mu  = \frac{E}{{2\left( {1 + \nu } \right)}}.
\end{equation}
\begin{figure}
\centering\includegraphics[width=0.8\linewidth]{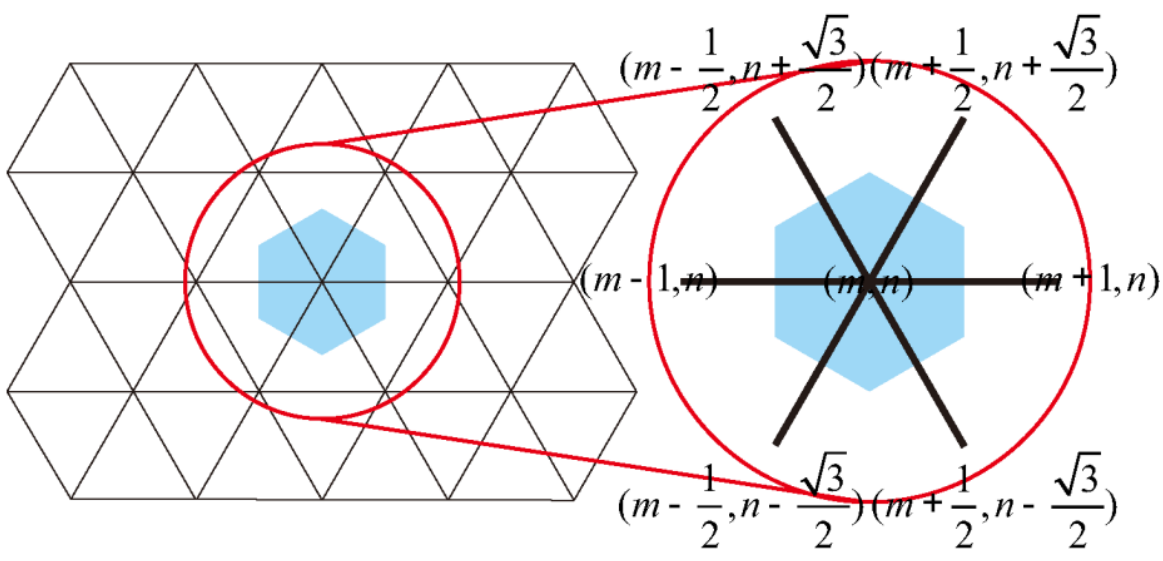}
\caption{Flowchart of lattice spring modeling}\label{fA-1}
\end{figure}

\section{Commonly Used Effective Medium Theories}
\setcounter{table}{0} 
\setcounter{figure}{0}
\setcounter{equation}{0}
\renewcommand{\thetable}{B\arabic{table}}
\renewcommand{\thefigure}{B\arabic{figure}}
\renewcommand{\theequation}{B\arabic{equation}}
\label{a3}
The anisotropic media generally described by the models of transverse isotropy (TI). Elasticity matrices, which present the components of the elasticity tensor, depend on the spatial orientation of TI models. Here, we start with the simplest fracture geometry, a single set of given orientation and parallel cracks, that has been extensively studied \cite{cheng1993crack,hudson1980overall,hudson1981wave}. In three dimensions, a TI elastic material is completely specified by five independent constants. The stiffness matrix is given by
\begin{equation}
\label{eq42}
\left( {\begin{array}{*{20}{c}}
{{C_{1111}}}&{{C_{1122}}}&{{C_{1133}}}&0&0&0\\
{{C_{1122}}}&{{C_{1111}}}&{{C_{1133}}}&0&0&0\\
{{C_{1133}}}&{{C_{1133}}}&{{C_{3333}}}&0&0&0\\
0&0&0&{{C_{2323}}}&0&0\\
0&0&0&0&{{C_{2323}}}&0\\
0&0&0&0&0&{{C_{1212}}}
\end{array}} \right),
\end{equation}
where ${C_{1212}} = \frac{1}{2}\left({{C_{1111}} - {C_{1122}}} \right)$. Hudson \cite{hudson1980overall,hudson1981wave} developed first- and second-order expansions for the effective stiffness of a crack-induced transversely isotropic medium by means of a scattering approach with small-aspect-ratio assumption. He gave the results as flowing,
\begin{equation}
\label{eq43}
C_{ijkl}^* = C_{ijkl}^0 + C_{ijkl}^1 + C_{ijkl}^2,
\end{equation}
$C_{ijkl}^*$, $C_{ijkl}^0$, $C_{ijkl}^1$, and $C_{ijkl}^2$ are the effective elastic stiffness, the matrix stiffness, the first- and second-order corrections. The first-order corrections are given by,
\begin{equation}
\label{eq44}
{C_{ijkl}}^1 =  - \frac{{C_{r3ij}^0C_{s3kl}^0}}{\mu }\varepsilon {U_{rs}}\left( 0 \right),
\end{equation}
and the second-order corrections are expressed by,
\begin{equation}
\label{eq45}
{C_{ijkl}}^2 = \frac{1}{\mu }C_{ijpq}^1C_{rskl}^1{\chi _{pqrs}},
\end{equation}
where for the dry cracks, the expressions are
\begin{equation}
\label{eq46}
{U_{11}}\left( 0 \right) = \frac{{16\left( {\lambda  + 2\mu } \right)}}{{3\left( {3\lambda  + 4\mu } \right)}},
\end{equation}
\begin{equation}
\label{eq47}
{U_{33}}\left( 0 \right) = \frac{{4\left( {\lambda  + 2\mu } \right)}}{{3\left( {\lambda  + \mu } \right)}},
\end{equation}
\begin{equation}
\label{eq48}
{U_{kl}}\left( 0 \right) = 0{\kern 1pt} {\kern 1pt} {\kern 1pt} {\kern 1pt} \left( {k \ne l} \right);
\end{equation}
\begin{equation}
\label{eq49}
{\chi _{ijkl}} = \frac{1}{{15}}\left( {{\delta _{ik}}{\delta _{jl}}\left( {4 + \frac{\mu }{{\lambda  + 2\mu }}} \right) - \left( {{\delta _{kj}}{\delta _{il}} + {\delta _{ij}}{\delta _{kl}}} \right)\left( {1 - \frac{\mu }{{\lambda  + 2\mu }}} \right)} \right),
\end{equation}
in which $\varepsilon $ is the crack density.

Cheng \cite{cheng1993crack} used the Pad\'e approximation to solve the divergent phenomenon when the expressions are in a power series as shown in Eq. (\ref{eq43}), namely,
\begin{equation}
\label{eq50}
C_{ijkl}^* = C_{ijkl}^0\frac{{1 - {a_{ijkl}}\varepsilon }}{{1 + {b_{ij}}\varepsilon }},
\end{equation}
where
\begin{equation}
\label{eq51}
{b_{ijkl}} =  - \frac{{C_{ijkl}^2}}{{C_{ijkl}^1\varepsilon }},
\end{equation}
\begin{equation}
\label{eq52}
{a_{ijkl}} =  - \frac{{C_{ijkl}^1}}{{C_{ijkl}^0\varepsilon }} - {b_{ijkl}}.
\end{equation}

\bibliographystyle{spmpsci}      

\end{document}